\documentclass[pdflatex,sn-mathphys-num]{sn-jnl}

\usepackage{graphicx}%
\usepackage{multirow}%
\usepackage{amsmath,amssymb,amsfonts}%
\usepackage{amsthm}%
\usepackage{mathrsfs}%
\usepackage[title]{appendix}%
\usepackage{xcolor}%
\usepackage{textcomp}%
\usepackage{manyfoot}%
\usepackage{booktabs}%
\usepackage{algorithm}%
\usepackage{algorithmicx}%
\usepackage{algpseudocode}%
\usepackage{listings}%
\usepackage{url}
\usepackage{pdflscape}    
\usepackage{adjustbox}    
\usepackage{booktabs}     
\usepackage{array}    
\usepackage{subfigure}
\usepackage{natbib}
\begin{document}

\title{Free Elections in the Free State:\break Ensemble Analysis of Redistricting in New Hampshire}

\author*[1]{\fnm{Atticus} \sur{McWhorter}}\email{atticus.w.mcwhorter.gr@dartmouth.edu}

\author[2]{\fnm{Daryl} \sur{DeFord}}

\affil[1]{\orgdiv{Department of Mathematics}, \orgname{Dartmouth College}}

\affil[2]{\orgdiv{Department of Mathematics and Statistics}, \orgname{Vassar College}}


\abstract{The process of legislative redistricting in New Hampshire, along with many other states across the country, was particularly contentious during the 2020 census cycle. In this paper we present an ensemble analysis of the enacted districts to provide mathematical context for claims made about these maps in litigation. Operationalizing the New Hampshire redistricting rules and algorithmically generating a large collection of districting plans allows us to construct a baseline for expected behavior of districting plans in the state and evaluate non-partisan justifications and geographic tradeoffs between districting criteria and partisan outcomes. In addition, our results demonstrate the impact of selection and aggregation of election data for analyzing partisan symmetry measures.}

\maketitle

\section{Introduction}\label{sec1}

While computational methods for optimizing political redistricting plans for multiple criteria have been been applied since at least the 1960's \cite{algs_chapter}, recent advances in data availability and computational power have created both new tools and new challenges for analyzing enacted and proposed maps. In this paper we apply one of these modern techniques, the ensemble method \cite{aftermath}, to create and analyze a large collection of districting plans for electoral districts in the State of New Hampshire. This approach allows us to formulate a baseline of reasonable values across a wide variety of metrics for each type of district, while directly incorporating the state's political geography and legislative constraints. 

Individual numerical measurements of compactness have long been applied to detect partisan gerrymandering, following the intuition that achieving improper partisan aims would require distorting and perturbing the boundaries of districts away from regular or smooth shapes. Examples of shape-based measurements include the isoperimetric ratio, known as the Polsby-Popper score in this domain, or the Reock score, which measures the ratio between a district and its circumscribing circle. These metrics have recently been criticized \cite{barnes_solomon_2021,jumble,duchin2018discrete} as insufficient to detect modern gerrymanders. North Carolina's Congressional maps in the previous cycle provide an example of this issue, as a particularly non-compact map was struck down as unconstitutional \cite{courtopinionNC1} and replaced with a map that had similar partisan properties but much better performance on the shape-based measures \cite{courtopinionNC2,Herschlag_NC}.

Similarly, many measurements have been proposed for directly evaluating the partisan fairness of a districting plan, without necessarily taking into account the shape of the districts. These include measures of partisan symmetry \cite{Grofman-1983,KKR}, as well as the efficiency gap \cite{bernstein2017formula,eg1,Veomett}, declination \cite{declination,warrington2017quantifying}, partisan dislocation \cite{eubank1, eubank2} and other recent geographic extensions \cite{BE, Wise}, and the GEO metric \cite{GEO}. These measures often imply specific and sometimes contradictory normative definitions of fairness with ideal values but relying on a single measurement of this sort to define a fair plan can sometimes lead to unintuitive results \cite{psymm, DV}.

Even simple measures of proportionality can fail to be achievable, even if it is not required or endorsed by the courts. For example, in several elections in Massachusetts with approximately a 30\% Republican vote share an outcome that does not elect any Republican representatives is required by a single member district system due to the homogeneity of the partisan distribution \cite{duchin2018locating}. That said, recent analysis of the ``Freedom to Vote Test'' with a proportional baseline \cite{DuchinSchoenbach} demonstrates possible advantages of this measure. 

The ensemble method attempts to address this problem by explicitly taking into account a specific state's political geography in the formation of districting plans. By creating a large collection of maps sampled from among the enormous collection of possibilities it is possible to provide context for the values of any metric applied to a specific map under consideration. For example, if the value of a metric applied to an enacted map lies in the extreme tail of the distribution of values over a neutral ensemble that provides evidence about the intentions of the line drawer \cite{Herschlag_NC}. 
Over the past decade, these ensemble methods have become increasingly relevant in court cases and reform efforts about political redistricting. State courts across the country 
have heard testimony about ensembles and many of the current Supreme Court justices has blessed their use in various applications. For example, in the Rucho v. Common Cause dissent, Justice Kagan argues that the creation of ensembles would prove beneficial toward determining whether a map is constitutional \cite{courtopinion1}, and as \cite{colorado} note, the methods have been used widely.

Ensemble methods have been widely successful both in analyzing the nation-level representative bodies and state legislatures \cite{caldera2020mathematics, carter2019optimal, deford2019redistricting, MO, Herschlag_NC, herschlag_evaluating_2017, zhao2022mathematically}. Especially in light of Rucho v. Common Cause, which ruled that partisan redistricting is not a justiciable issue, state-level courts have come to play an increasingly powerful role in the process of evaluating legislative maps, leading to increased attention.

In the recent court challenge to the New Hampshire's State Senate and Executive Council districts, no ensemble analysis was used to support evaluations of quantitative measures of symmetry, compactness, or fairness  applied to the enacted plans \cite{complaintNH}.  The plaintiffs do argue that misshapen districts were formed to pack together left-leaning cities to increase Republican representation in the remainder of the state. Additionally, the lawsuit does not provide a complete analysis of the impacts of the volatility of the electorate on the fairness metrics. As we demonstrate in Figure \ref{fig:nh_votes}, 
in a single year voters may overwhelmingly support a Democratic candidate for President and a Republican candidate for Governor. And as such, the choice of election data has large ramifications on the election outcome when used to analyze districting plans. 

To analyze this issue and provide additional content for the evaluation of the plans, we  apply quantitative measures to an ensemble of millions of plausible State Senate and Executive Council districting plans. These ensembles are overlaid with data from statewide elections that occurred during the previous census cycle, allowing us to perform a detailed analysis of the currently enacted plan. We then proceed to address the claims made in the lawsuit and other related evaluations of the potential maps for the state. 

\section{Methods}\label{sec3}

This section describes the technical choices that generated the ensemble. Most computations were done on an Apple M2 CPU, with memory-intense tasks being preformed on Dartmouth's {\texttt doob} cluster, a 96-core 3.6GHz AMD Epyc system. The project uses the 0.2.17 release of GerryChain. 

\subsection{Data}

The data for this project was taken from the Metric Geometry and Gerrymandering Group (MGGG) github. Precinct geometries are from NH Granit, New Hampshire's statewide GIS Clearinghouse, voting data is from the NH Secretary of State's Election Results page, and population data was obtained from the census API. All of these were already packaged together in shapefiles on the MGGG github page. For this project, we updated the shapefiles with the current State Senate and Executive Council district shapefiles from NH Granit. 

\begin{figure}[!htbp]
    \centering
    \includegraphics[width=.9\textwidth]{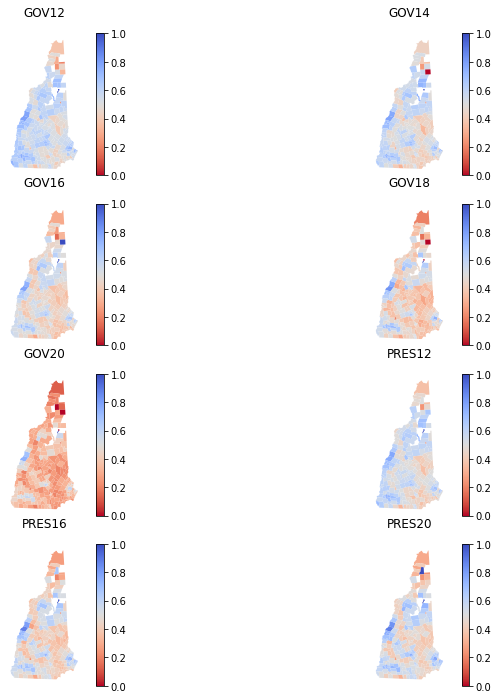}
    \caption{Voteshares in each precinct for the elections that we analyze in this paper. }
    \label{fig:nh_votes}
\end{figure}

After an ensemble of allowable districting plans is created, voting data is used to investigate partisanship. It has become good practice in computational redistricting papers to use state-wide elections as predictors of how voters will vote in local elections \cite{colorado, deford2019redistricting}. But in New Hampshire, split-ticket elections are more common, so it becomes difficult to choose which election data to use. In the 2020 governor's race, Republican Chris Sununu defeated Democrat Dan Feltes with a margin of over 30\% while in the presidential race, Joe Biden defeated Donald Trump with a 7\% margin. In order to address this issue, this paper will investigate the same ensembles through the lens of 8 different elections, the 2012, 2016, and 2020 presidential elections and the 2012, 2014, 2016, 2018, and 2020 gubernatorial elections, which we will refer to as PRES12, PRES16, PRES20, GOV12, GOV14, GOV16, GOV18, and GOV20, respectively.

\subsection{Markov Chain Monte Carlo Methods}
To create an ensemble, first the state dual graph must be created. Each node in the graph represents a precinct, and adjacent precincts are connected with an edge. Each node is weighted according to its population, and thus the problem of redistricting is reduced to the problem of graph partitioning under the constraint that each part should have equal weight. 

Ensembles were generated with Recombination (RECOM), an algorithm that generates a Markov chain of partitions. At each iteration, two districts are combined, and each edge is weighted randomly. Then, Kruskal's algorithm is used to find a maximum weight spanning tree, and a random edge is cut such that each new district satisfies the correct population deviation. For a more in depth discussion of the algorithm, see \cite{DeFord2021Recombination}. In this paper, the maximum allowed population deviation was set at 5\% for the State Senate and 1\% for the Executive Council. 

\begin{figure}[!htbp]
    \centering
    \includegraphics[scale=0.5]{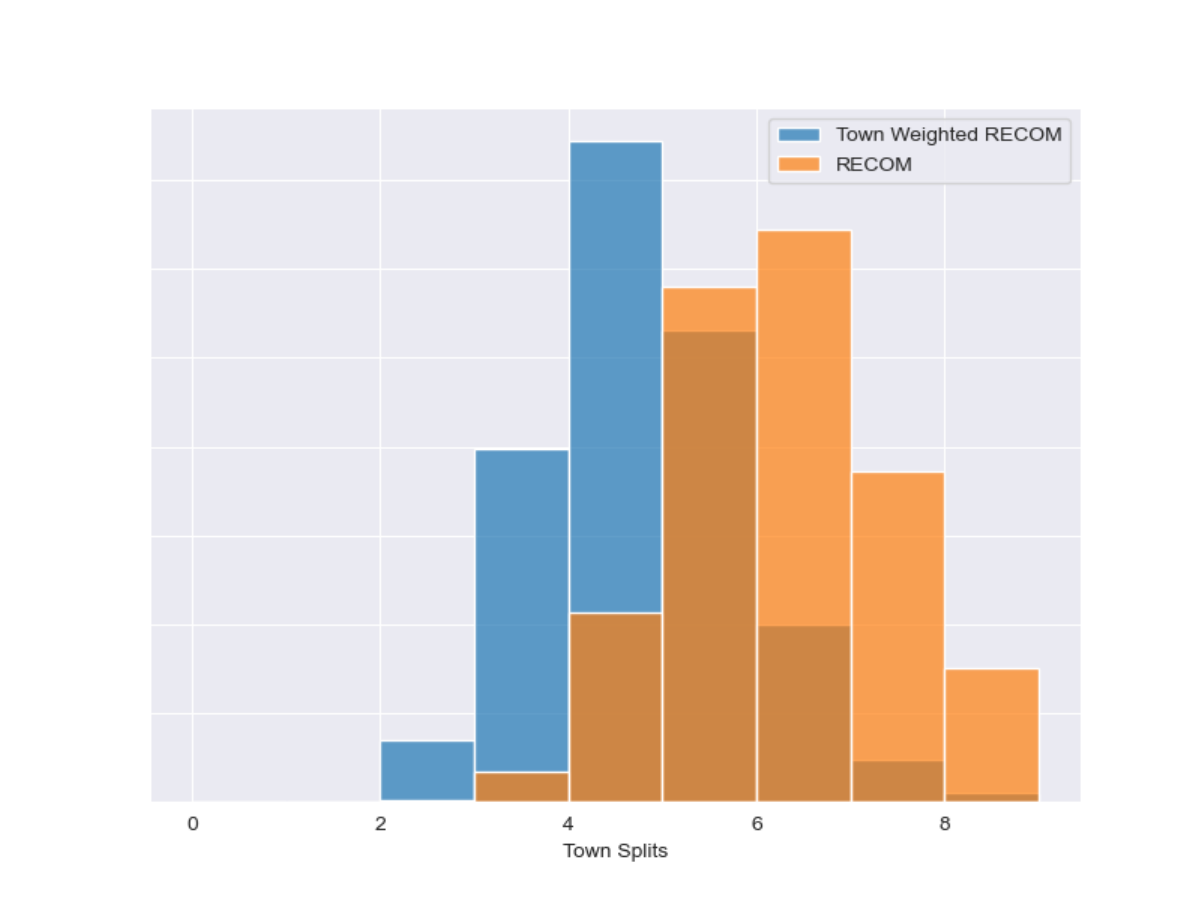}
    \caption{Number of split towns in a town-weighted and not town-weighted chain. The current enacted plan has three split towns. }
    \label{fig:town_weighting}
\end{figure}

New Hampshire requires districts to be comprised of contiguous towns, cities, and wards without splitting them. Thus we employ town weighting; weights for edges that connect nodes in the same town are drawn from the uniform distribution on [0, 4], and if the precincts are not in the same town, the edge weights are drawn from the uniform distribution on [0, 1]. The town weighted variant of RECOM was originally proposed for county weighting in \cite{colorado}, where the authors observe that county weighting reduces county splits to a realistic level. The upper bound on the uniform distribution was chosen to balance mixing time (town weighting adds rigidity to the chain) and reduction of split towns. In Figure \ref{fig:town_weighting}, a plot of the number of split towns is shown for two ensembles, one with and one without town-weighting. 

To set the length of the Markov chains, we relied on two heuristics of mixing. Firstly, to investigate mixing time of a single chain, autocovariances were calculated. If a chain is sufficiently long, then a good choice of stopping time is a large multiple of the time it takes for the autocovariance to decay to nearly zero. The second heuristic investigated is the multi-start heuristic; if multiple runs of the chain with different starting locations converge to the same distribution, then it is concluded that the chains have been run long enough. There are many ways to measure the distance between distributions, but in this paper, we employ the Kolmogorov-Smirnov (KS) metric as suggested in \cite{colorado}. The KS metric is equivalent to the infinity-norm on the cumulative distribution functions of the two samples. Thus, empirical distributions are calculated for each chain, and the maximum difference in the y-direction between two empirical distributions is the KS distance. 

For the autocovariance heuristic, a very long sample (20 million steps for the State Senate and 5 million steps for the Executive Council) is taken, and autocovariance is calculated as a function of lag-time.  For the multi-start heuristic, ten seeds were created using the bipartition-tree algorithm outlined by \cite{DeFord2021Recombination}, and the enacted plan was included as the eleventh seed. Then, from each seed chains are run (for 10 million steps for the State Senate and for 500,000 steps for the Executive Council), and KS distances are calculated for each pair of ensembles. The average pairwise KS distances and the lags where autocovariance drops below 0.01 are included in Tables 1-4 of the appendix.

\section{Results and Analysis}\label{sec4}
In this section we describe the results of our ensemble analysis across a wide variety of elections from the previous census cycle, corresponding to data that would have been available to the line drawers when the plan was created. 
\subsection{Executive Council}

We begin our analysis with the Executive Council, which divides the state into five districts.  We first compare the compactness of the ensemble to the enacted plan before evaluating the partisan properties. 

\subsubsection{Compactness Measures}

\begin{figure}[!htbp]
    \centering
    \subfigure[]{\includegraphics[width=0.49\textwidth]{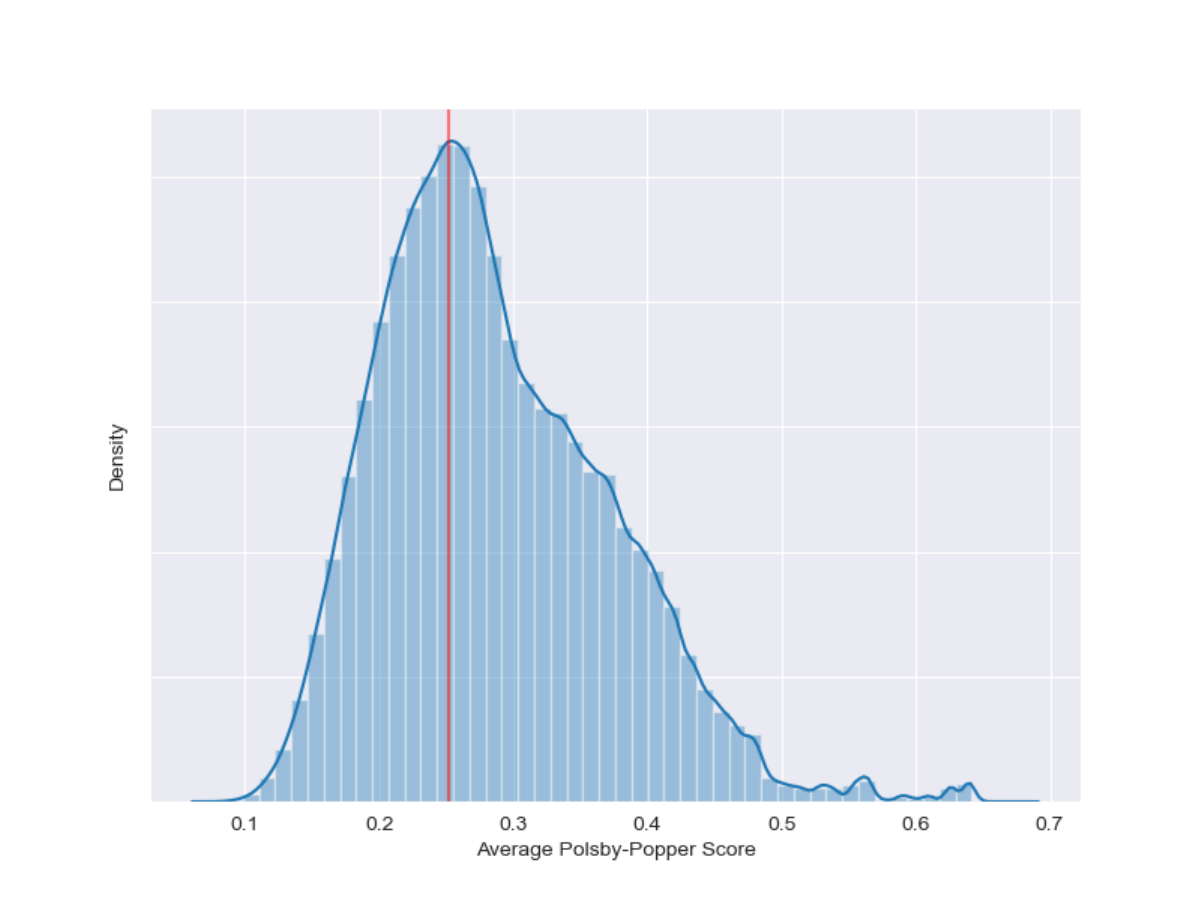}}
    \subfigure[]{\includegraphics[width=0.49\textwidth]{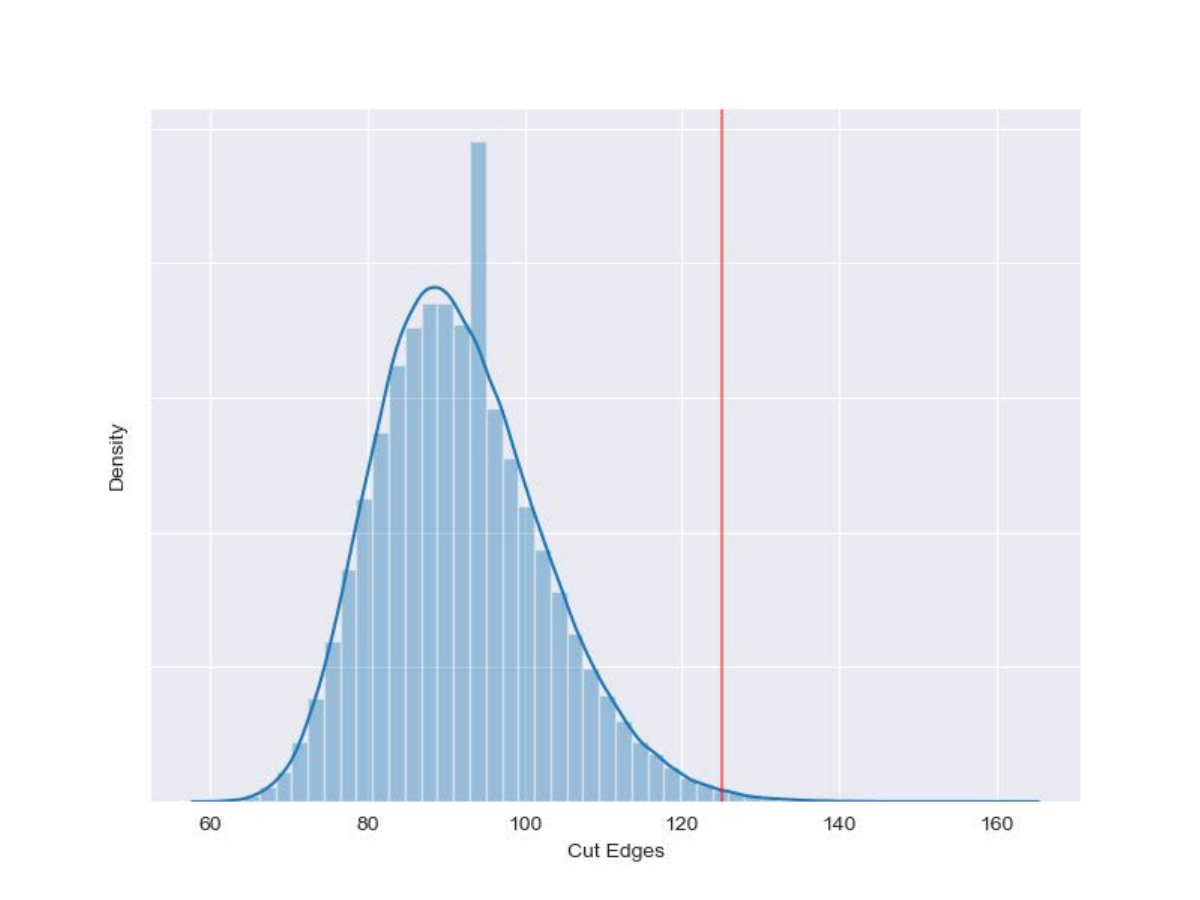}}
    \caption{(a) Average Polsby-Popper score of Executive Council districts (b) Number of cut edges in the graph of Executive Council districts}
    \label{fig:ec_compactness}
\end{figure}

We analyze two metrics of compactness: the Polsby-Popper score and the number of cut edges. The Polsby-Popper score measures the ratio of a shape's area to that of a circle with the same perimeter. It has some noted problems \cite{deford2019total}, but it has been widely accepted among political scientists \cite{moncrief2011reapportionment, Herschlag_NC}. Cut edges are a measurement proxy for compactness that are convenient for spanning-tree methods. For a more detailed exploration of cut edges as a proxy for compactness, see \cite{duchin2022explainer}. 

Figure \ref{fig:ec_compactness} show the ensemble's Polsby-Popper scores and cut edges. The enacted plan lies right on the ensemble mode for average Polsby-Popper score, but has many more cut edges. Thus the ensemble is at least as compact as the enacted plan.

\subsubsection{Seats and Win Percentages}
New Hampshire sees a large swing in election outcomes (see Figure \ref{fig:nh_votes}) even between races on the same ballot. The candidate up for election appears to have a greater effect than in other states, where split-ticket results are less common. Figure \ref{fig:nh_votes} shows the vote shares in each of the statewide elections that we analyze. This variability makes it difficult to draw meaningful conclusions from a single election. We follow \cite{firewalls} by investigating all recent statewide elections and searching for `firewalls', patterns that may appear symmetric, but encode biased results in close elections. Additionally, we investigate and interpret measurements of partisan symmetry of the plans. The dominant result is that the enacted plan is consistently more Republican leaning than the ensemble.

\begin{figure}[!htbp]
    \centering
    \subfigure[]{\includegraphics[width=0.49\linewidth]{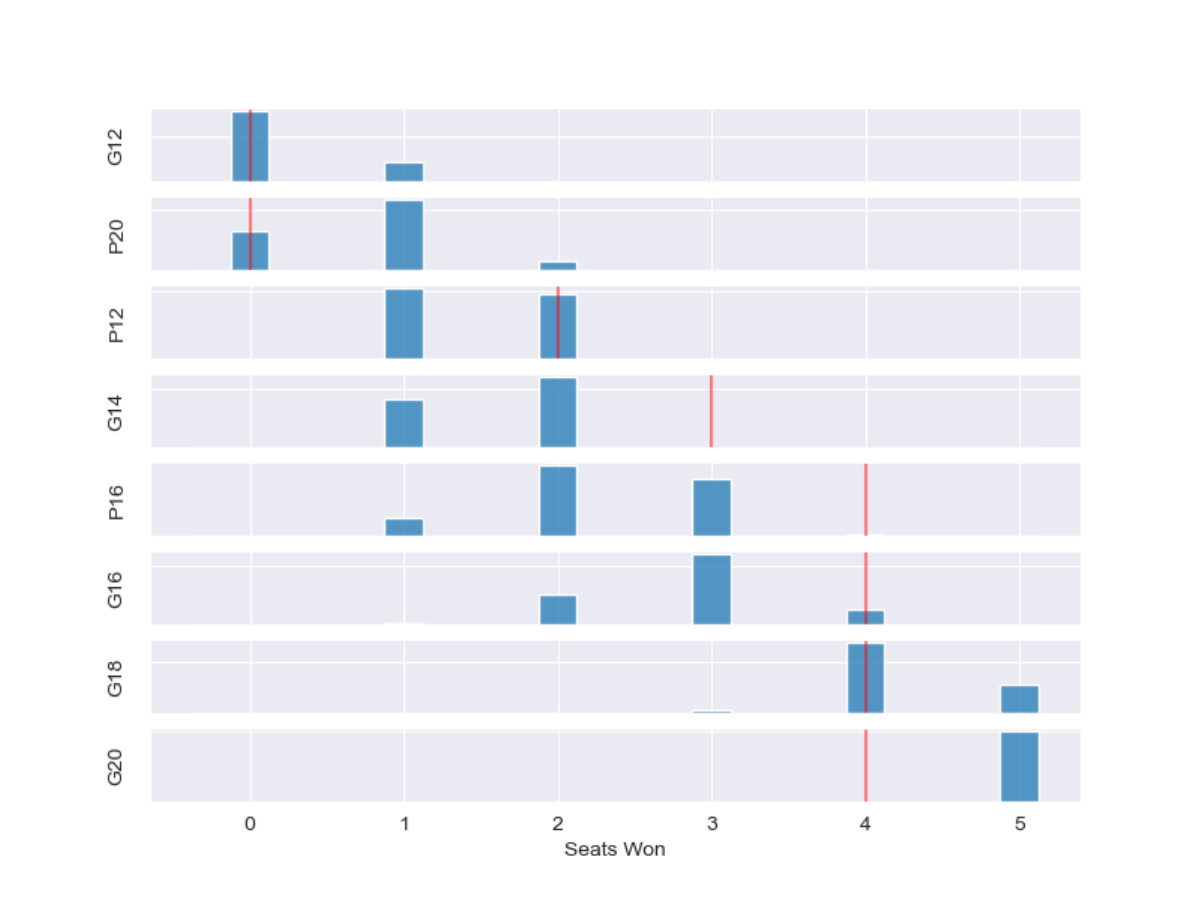}}
    \subfigure[]{\includegraphics[width=0.49\linewidth]{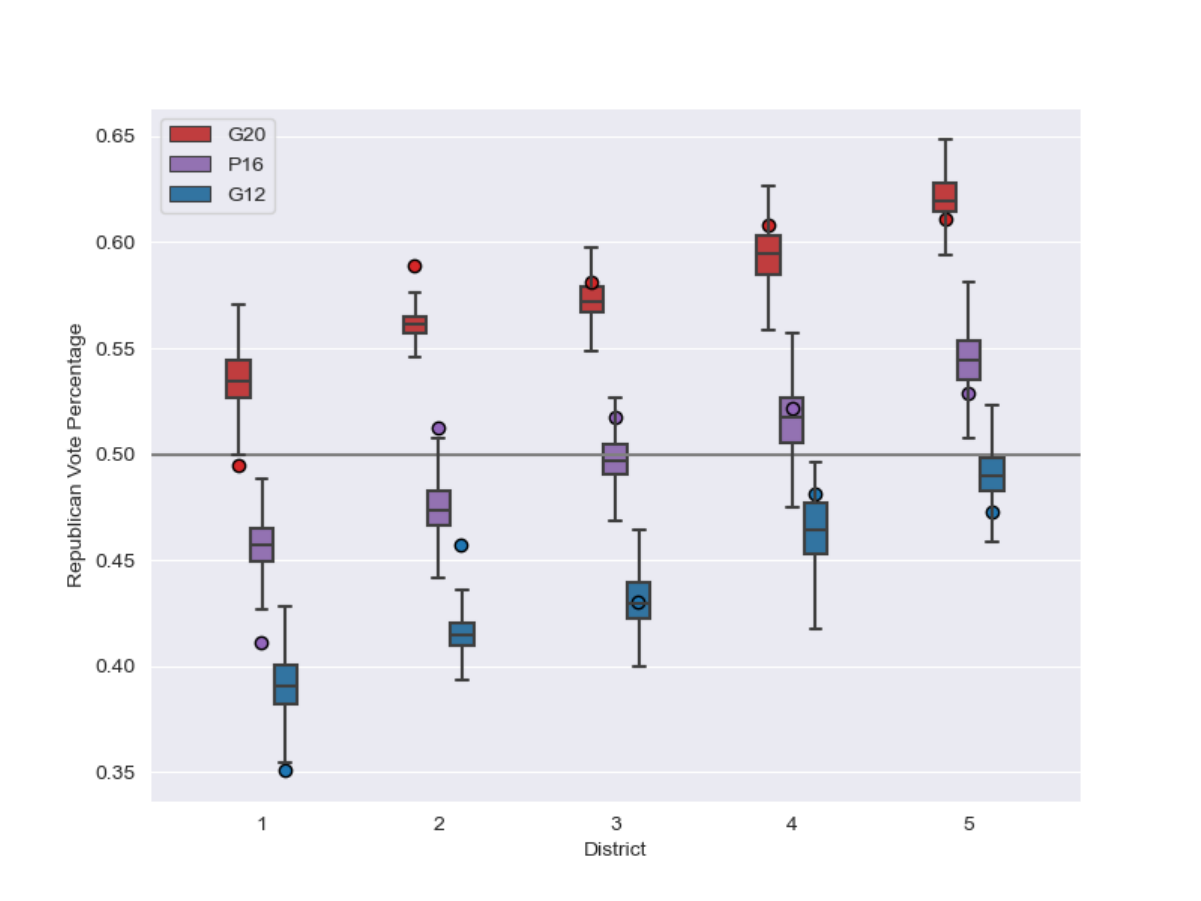}}
    \caption{(a) Executive Council Seats Won (b) Executive Council Republican win percentages sorted by district}
    \label{fig:Seats_Wins}
\end{figure}

In Figure \ref{fig:Seats_Wins}(a) we show the seats won by the Republican party. In elections with large win-margins, the enacted plan (red) lies on or to the left of the ensemble (blue), indicating an advantage for Democrats. In closer elections, the enacted plan lies to the right of the ensemble, indicating an advantage for Republicans. This could be an early indication of a Republican firewall. 

In figure \ref{fig:Seats_Wins}(b) we plot the distributions of the sorted vote totals in each district for 3 representative elections, as in \cite{caldera2020mathematics, colorado, deford2019redistricting, Herschlag_NC, herschlag_evaluating_2017, zhao2022mathematically}.
In the districts with the largest and smallest Republican vote shares, Republicans are underrepresented, and in the middle districts, Republicans are overrepresented. In a tight election such as PRES16, this overrepresentation gives Republicans an advantage.

\subsubsection{Partisan Symmetry Measures}
In this paper we investigate the mean-median score, efficiency gap, and partisan bias. For a more detailed discussion of partisan symmetry measures, see \cite{bernstein2022measuring}

\begin{figure}[!htbp]
    \centering
    \includegraphics[scale=0.5]{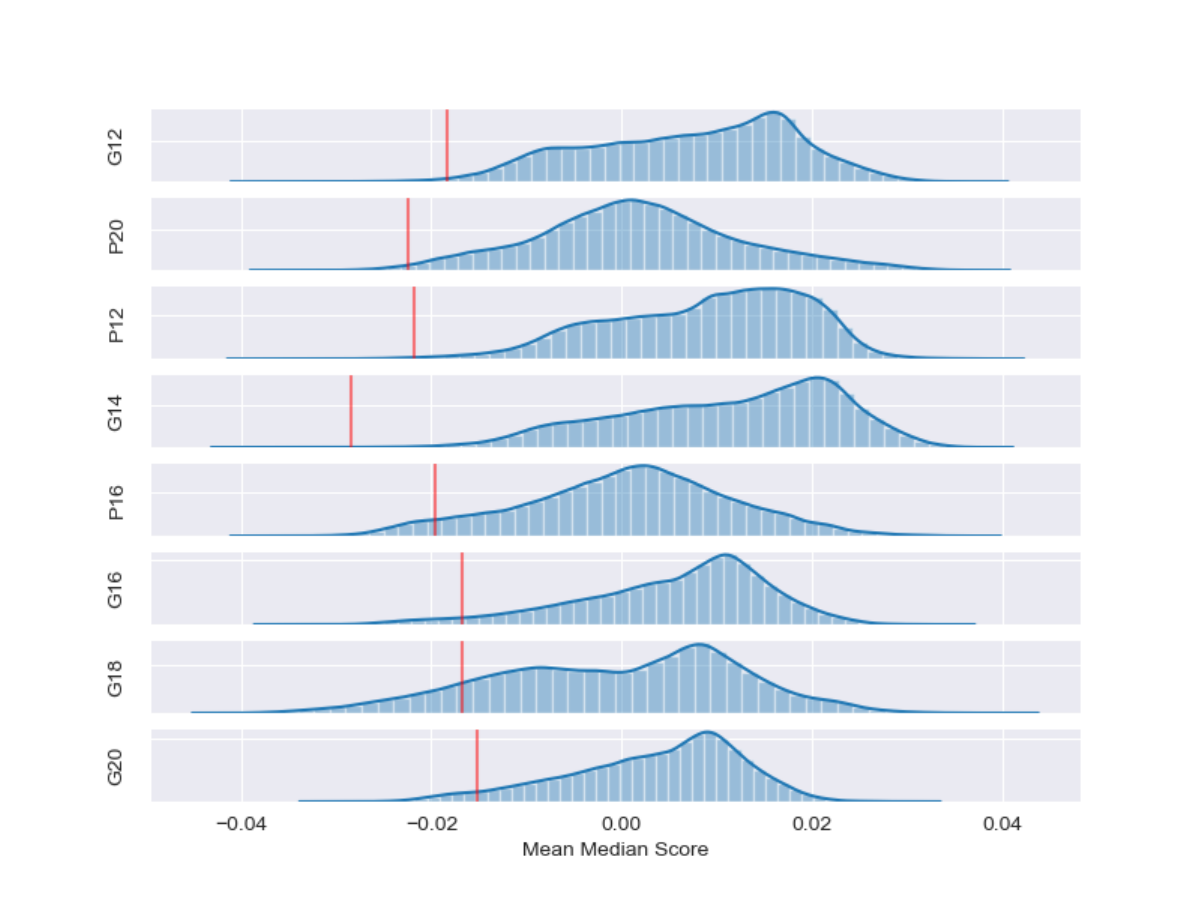}
    \caption{Executive Council Mean Median Scores across Elections. The red line denotes the value of the enacted plan, which has a Republican-favoring value in each studied election.  }
    \label{fig:ec_mm}
\end{figure}

The mean-median score measures a party's mean vote share minus its median vote share across districts. A positive score suggests the party may face a disadvantage. In our analysis, a positive mean-median score for Republicans indicates a Democratic advantage, while a negative score indicates a Republican advantage. Mean-median scores are plotted in Figure \ref{fig:ec_mm}. Regardless of election, the mean-median score for the enacted plan is negative, indicating a Republican advantage. On the other hand, the ensemble mean-median scores are positively centered, indicating a slight Democratic advantage. In the case of mean-median score, New Hampshire's political geometry encodes a slight Democratic advantage, but despite this, the enacted plan encodes a clear Republican advantage. 

\begin{figure}[!htbp]
    \centering
    \includegraphics[scale=0.5]{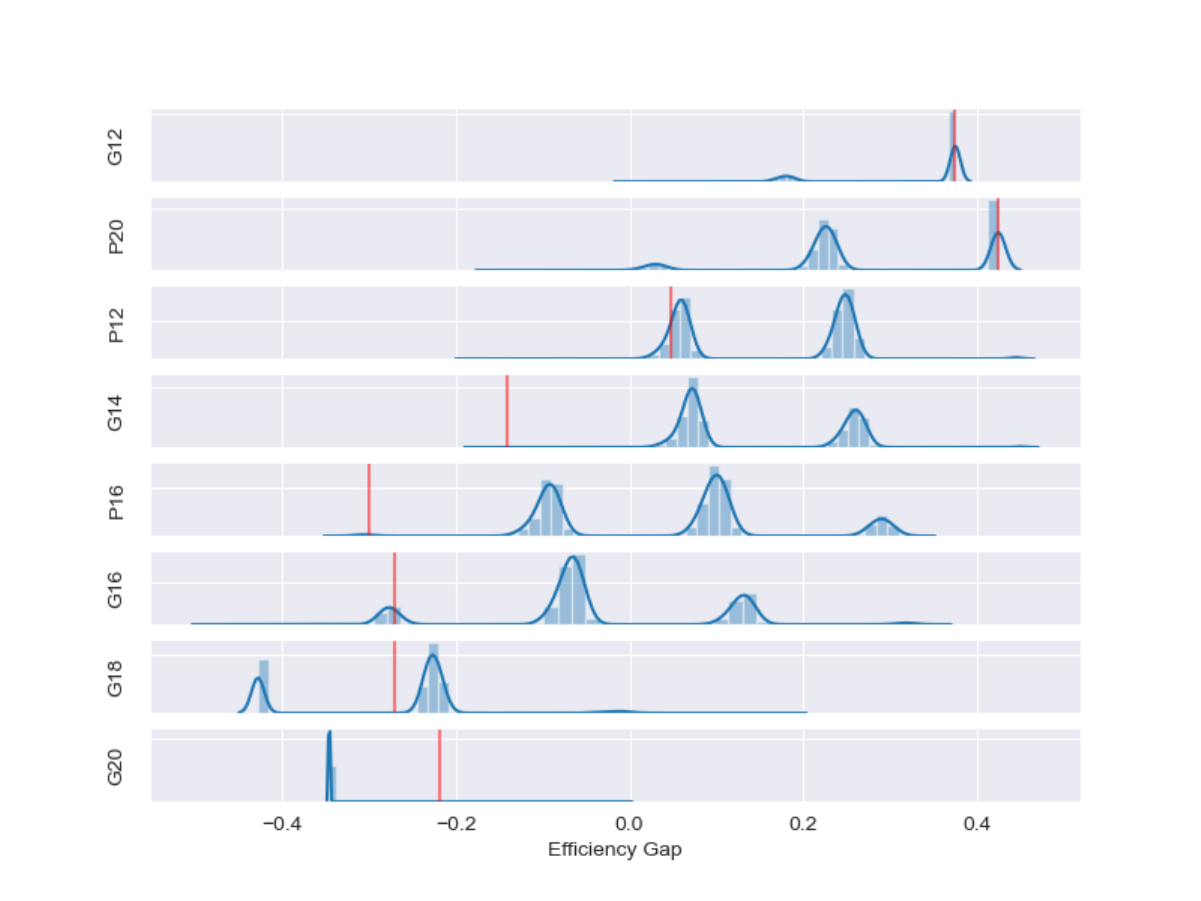}
    \caption{Executive Council Efficiency Gap Across Elections. The red line denotes the value of the enacted plan.}
    \label{fig:ec_eg}
\end{figure}

The Efficiency Gap measures the difference between parties' wasted votes. The efficiency gap values are shown in Figure \ref{fig:ec_eg}, where a negative value indicates a Republican advantage and a positive value indicates a Democratic advantage. In large wins, the enacted plan lies right of the ensemble, indicating an advantage for Democrats, but in close elections, the enacted plan encodes an advantage for Republicans. 

\begin{figure}[!htbp]
    \centering
    \includegraphics[scale=0.5]{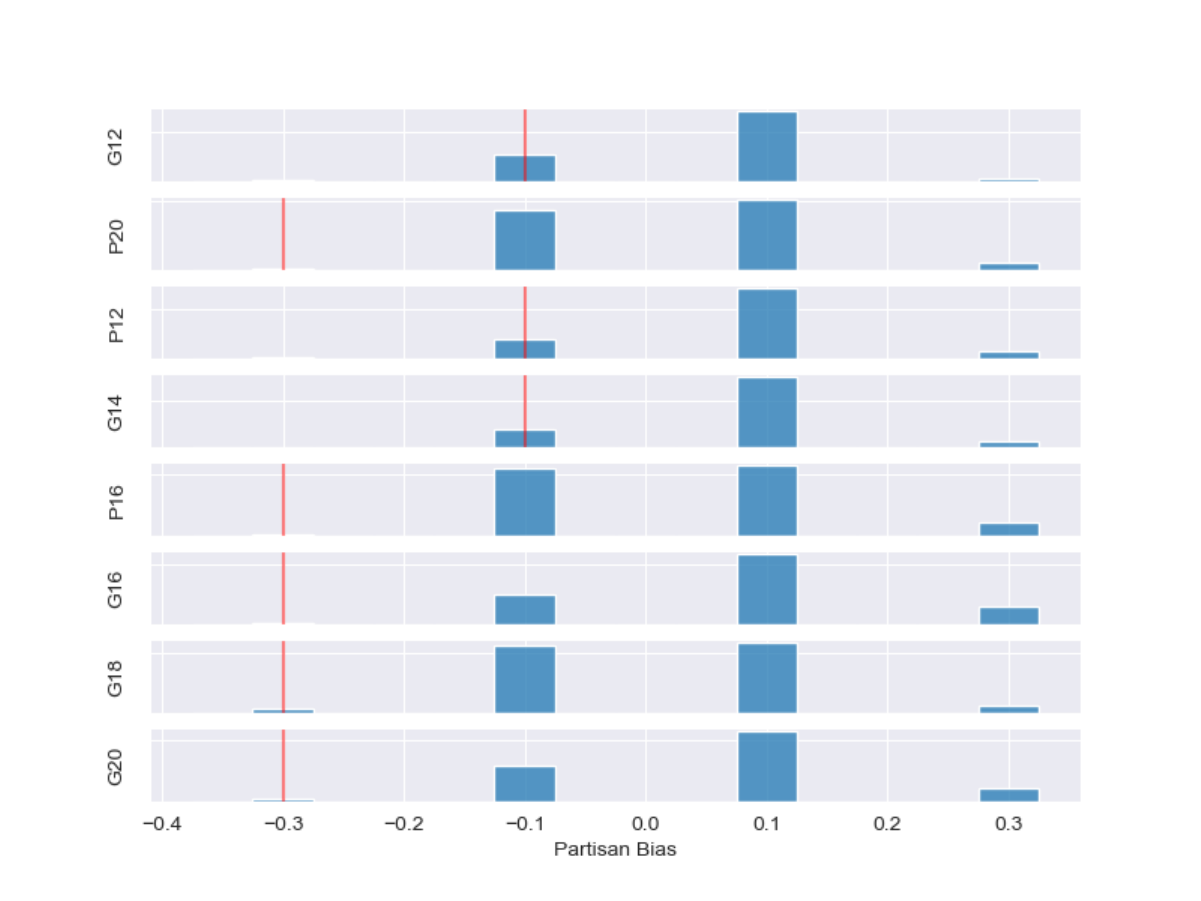}
    \caption{Executive Council Partisan Bias across Elections}
    \label{fig:ec_pb}
\end{figure}

The partisan bias score reflects a party's advantage in a hypothetical 50-50 election. The partisan bias score increases or decreases in steps inversely proportional to the number of seats, resulting in discrete ensemble distributions. A positive score indicates a Democratic advantage, while a negative score indicates a Republican advantage. In Figure \ref{fig:ec_pb}, the ensemble mode typically occurs around 0.1, signaling a tendency for plans to encode a Democratic advantage. However, the enacted plan favors Republicans.

Analysis of partisan symmetry measures supports the conclusion that the enacted plan favors Republicans more than the ensemble.

\subsubsection{A Close Look at the Districts}

\begin{figure}[!htbp]
    \centering
    \subfigure[]{\includegraphics[width=0.32\linewidth]{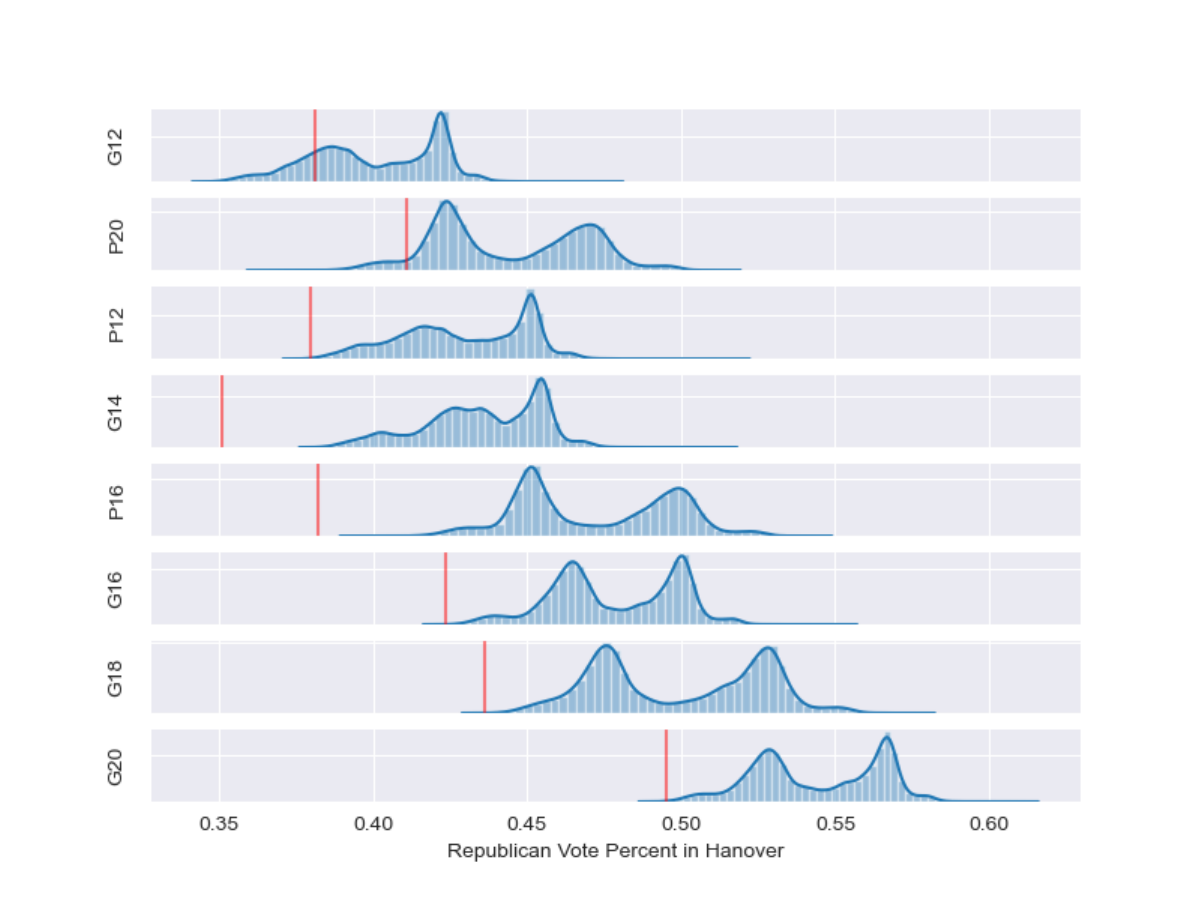}}
    \subfigure[]{\includegraphics[width=0.32\linewidth]{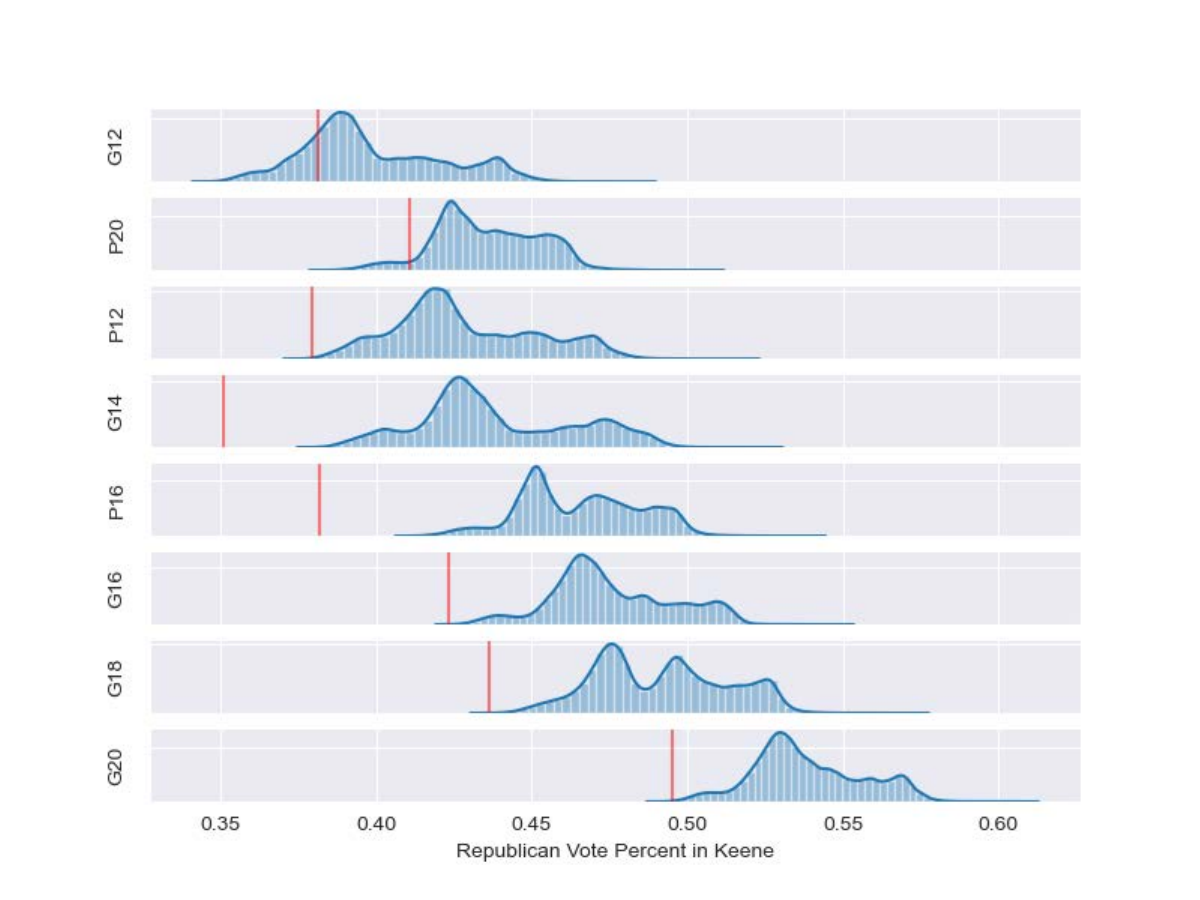}}
    \subfigure[]{\includegraphics[width=0.32\linewidth]{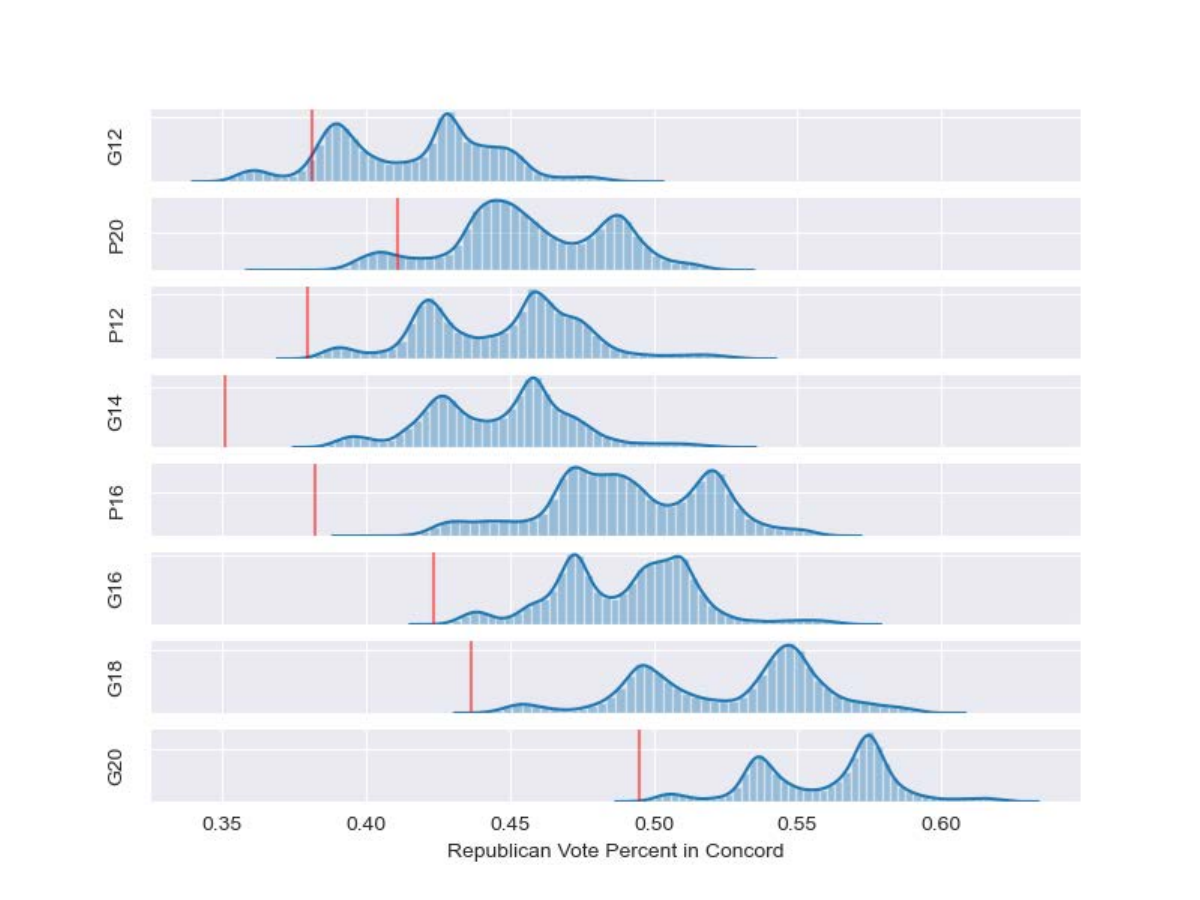}}
    \subfigure[]{\includegraphics[width=0.32\linewidth]{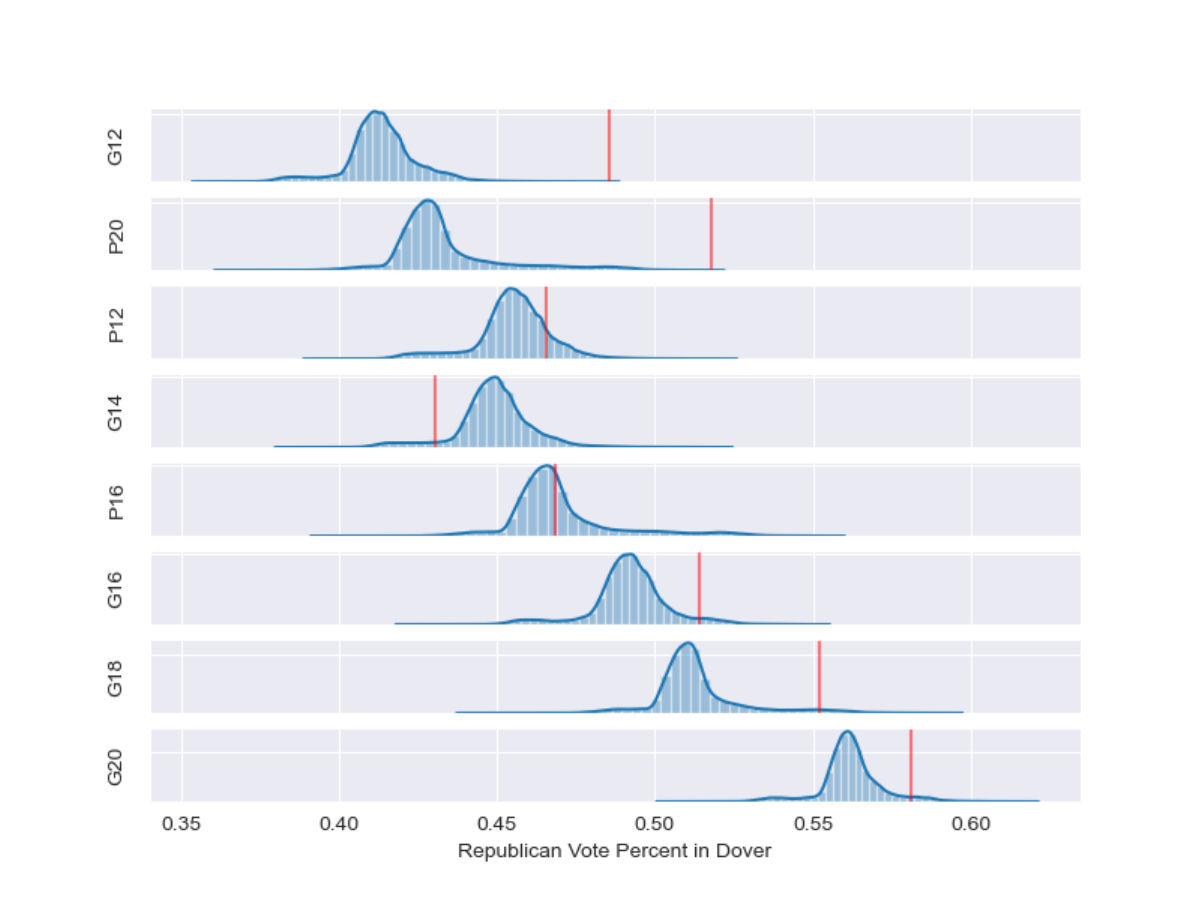}}
    \subfigure[]{\includegraphics[width=0.32\linewidth]{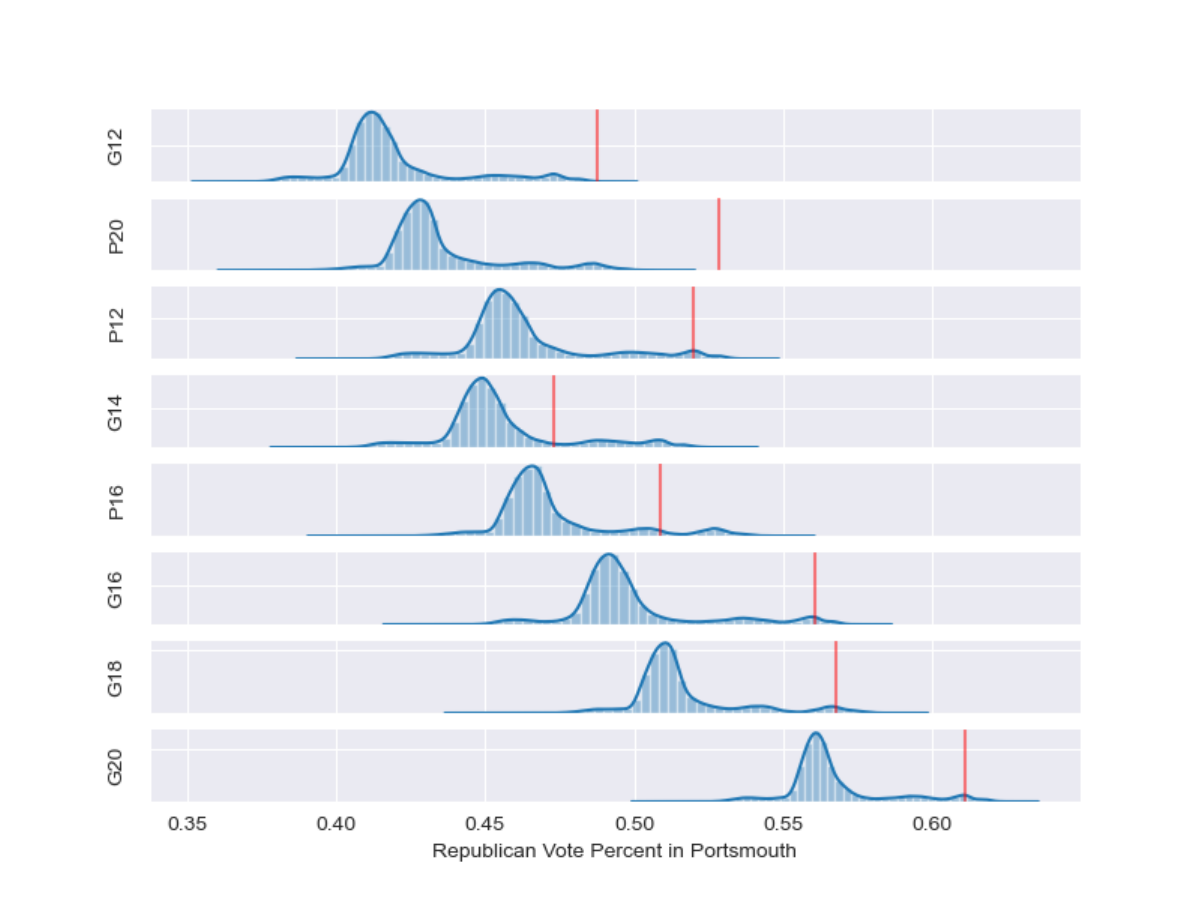}}
    \caption{Republican vote percentages in the districts containing (a) Hanover (b) Keene (c) Concord (d) Dover (e) Portsmouth}
    \label{fig:EC_towns}
\end{figure}

Analysis of the vote totals indicates that district two, which includes population centers like Hanover, Keene, and Concord, is packed with Democrats. Figure \ref{fig:EC_towns}(a)-(c) illustrates the vote totals for these towns, showing that in every election, the Republican vote percentage is significantly lower than expected. For instance, in GOV20, despite Republican candidate Sununu winning the popular vote by over 30\%, a Democrat still carried the district. This explains why the efficiency gap for this election favors Democrats; even with substantial Republican victories, the packed district allows Democrats to secure a seat.

Other representative districts are those containing Portsmouth and Dover, whose vote totals are plotted in Figures \ref{fig:EC_towns}(d) and \ref{fig:EC_towns}(e). Here, the trend is opposite: in Portsmouth the enacted plan gives Republicans a higher vote than the ensemble mode and Dover's district follows a similar pattern.

This analysis supports the claims made in \cite{complaintNH}. The lawyers argue that the population centers of Concord and Keene are grouped unnaturally with the left leaning western border. Our findings show that Republican vote totals in Hanover, Keene, and Concord in the enacted plan are much lower than those in the ensemble. This grouping takes Democratic support away from other districts, evident in Dover and Portsmouth, where the enacted plan pushes Republican vote totals above 50\%.

Another observation is the bimodal distribution of the vote totals. Hanover, being near the Vermont border, can be districted into the more Republican Northern region (as in the 2010 plan) or the more Democrat southeastern region (as in the 2020 plan). The same reasoning can be applied to Concord. Even taking this multimodality into account, the enacted plan still appears to be an outlier, suggesting that the choice of direction to extend the district which contains Hanover is not sufficient to explain the full extent of the deviation from the ensemble. 

\subsection{State Senate}

We also performed a similar analysis for the New Hampshire State Senate, which has 24 districts. 

\subsubsection{Compactness Measures}
\begin{figure}[!htbp]
    \centering
    \subfigure[]{\includegraphics[width=0.49\linewidth]{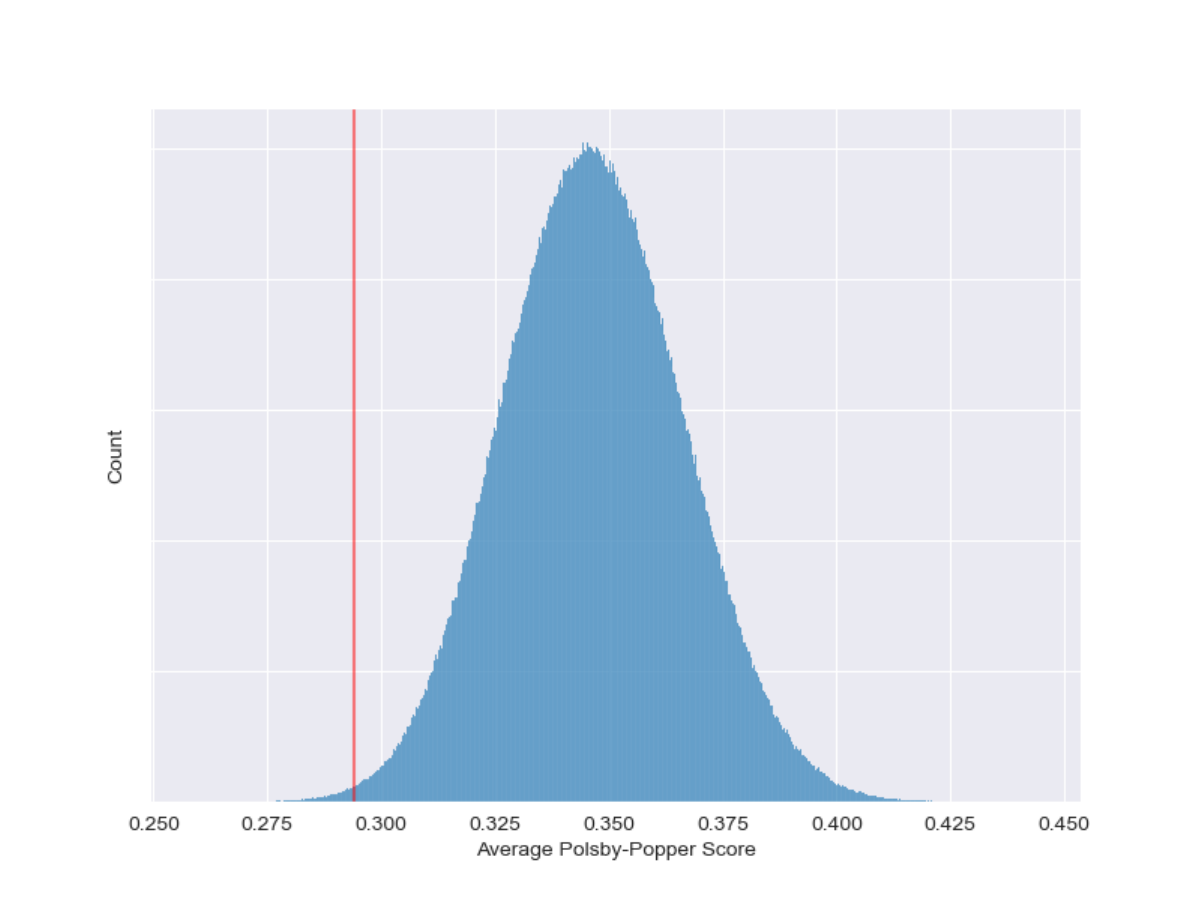}}
    \subfigure[]{\includegraphics[width=0.49\linewidth]{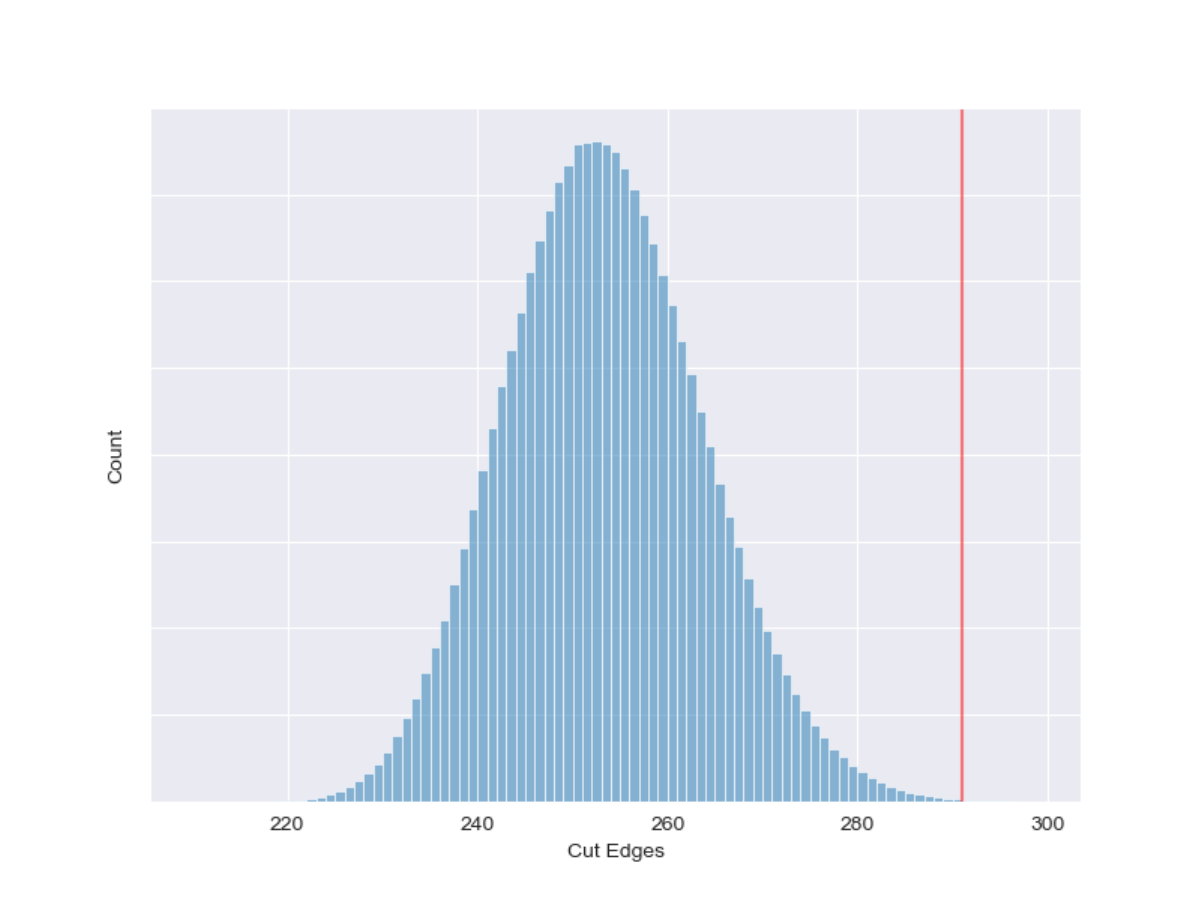}}
    \caption{(a) Average Polsby-Poppy Score of State Senate Districts (b) Number of Cut Edges}
    \label{fig:ss-comp}
\end{figure}

As is evident in Figure \ref{fig:ss-comp}, the relation between the compactness of the ensemble and the enacted plan depends on the chosen metric; The ensemble has fewer cut edges and higher average Polsby-Popper scores than the enacted plan.

\subsubsection{Seats and Win Percentages}
\begin{figure}[!htbp]
    \centering
    \subfigure[]{\includegraphics[width=0.49\linewidth]{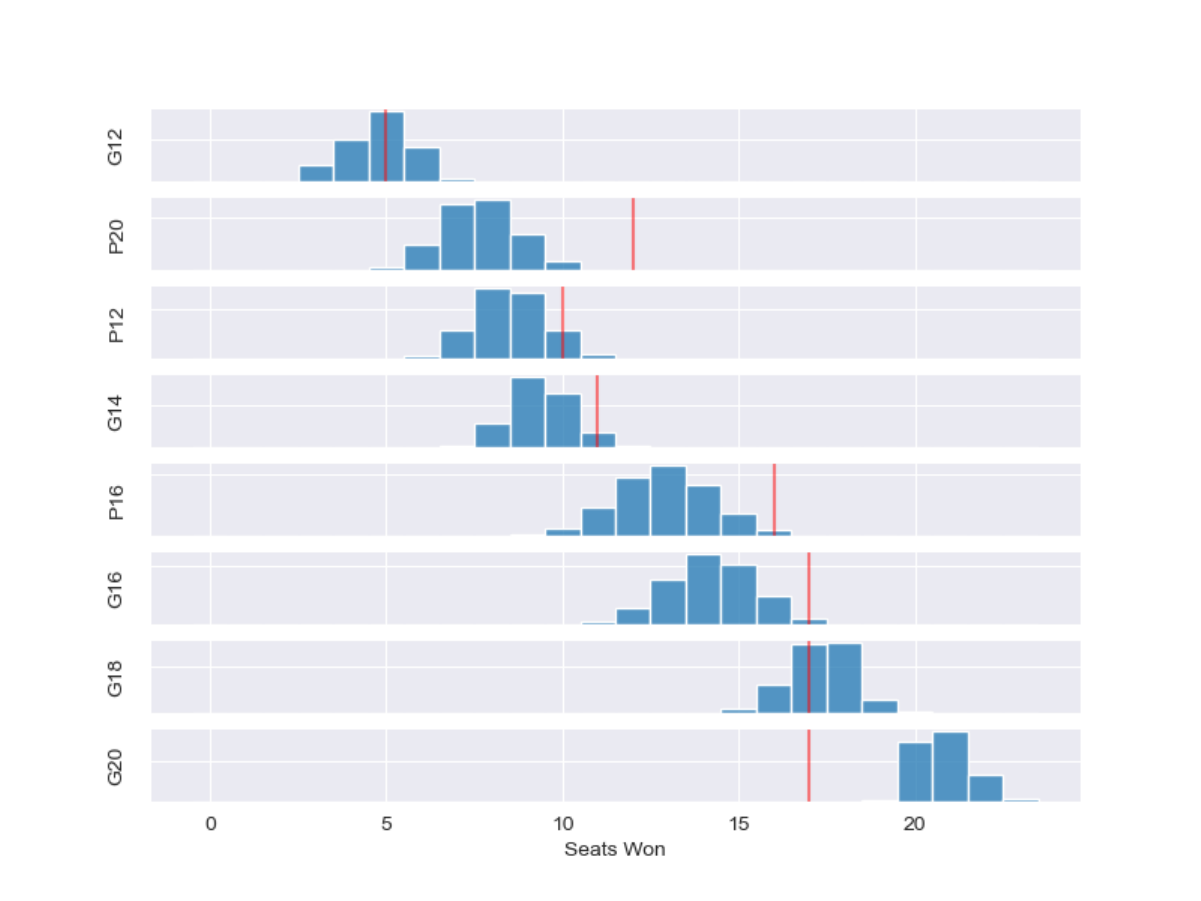}}
    \subfigure[]{\includegraphics[width=0.49\linewidth]{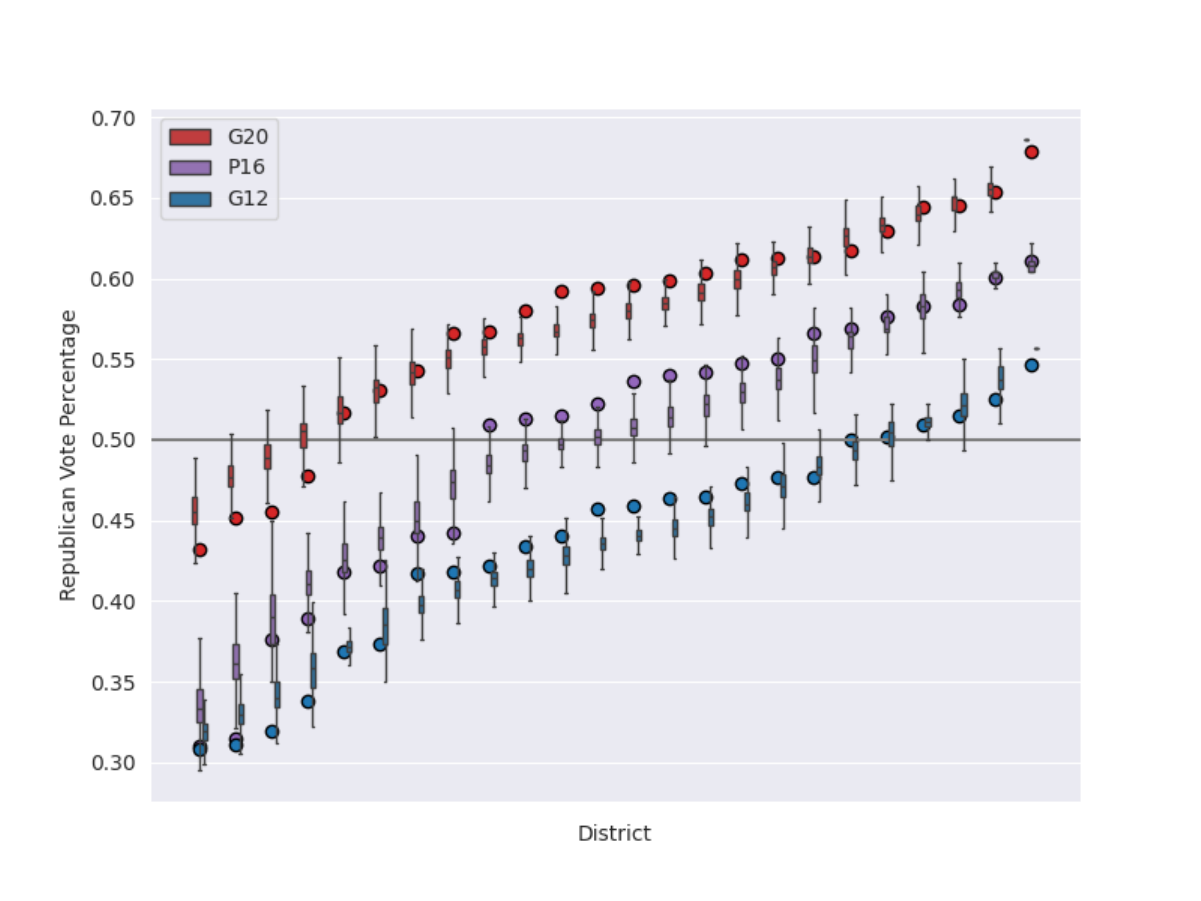}}
    \caption{(a) State Senate Seats Won (b) State Senate Republican win percentages sorted by district}
    \label{fig:ss-seat}
\end{figure}

In Figure \ref{fig:ss-seat}(a), we observe similar behavior to the executive council. In elections with large win-margins, the enacted plan gives an advantage to Democrats, and in close elections, the enacted plan gives an advantage to Republicans. 

The vote totals of each district are plotted in Figure \ref{fig:ss-seat}(b). In the most Republican and least Republican districts, the Republican vote total of the ensemble is lower than that of the enacted plan, but in the middle, the enacted plan lies above the ensemble. In landslide elections like GOV12 and GOV20, this doesn't encode a large trade-off of seats, as the middle districts are not highly contested. However, in tight elections like PRES16, the middle districts lie around 50\%, and as such, over-representation in those districts translates directly to over-representation in the number of seats won. 

This points to an asymmetry in the enacted plan; when Republicans have a narrow victory in the statewide vote (GOV16), the enacted plan yields a large victory in seats, but when Democrats have a narrow victory in the statewide vote (PRES16 or PRES20), Republicans maintain a majority of seats.

\subsubsection{Partisan Symmetry Measures}
\begin{figure}[!htbp]
    \centering
    \subfigure[]{\includegraphics[width=0.49\linewidth]{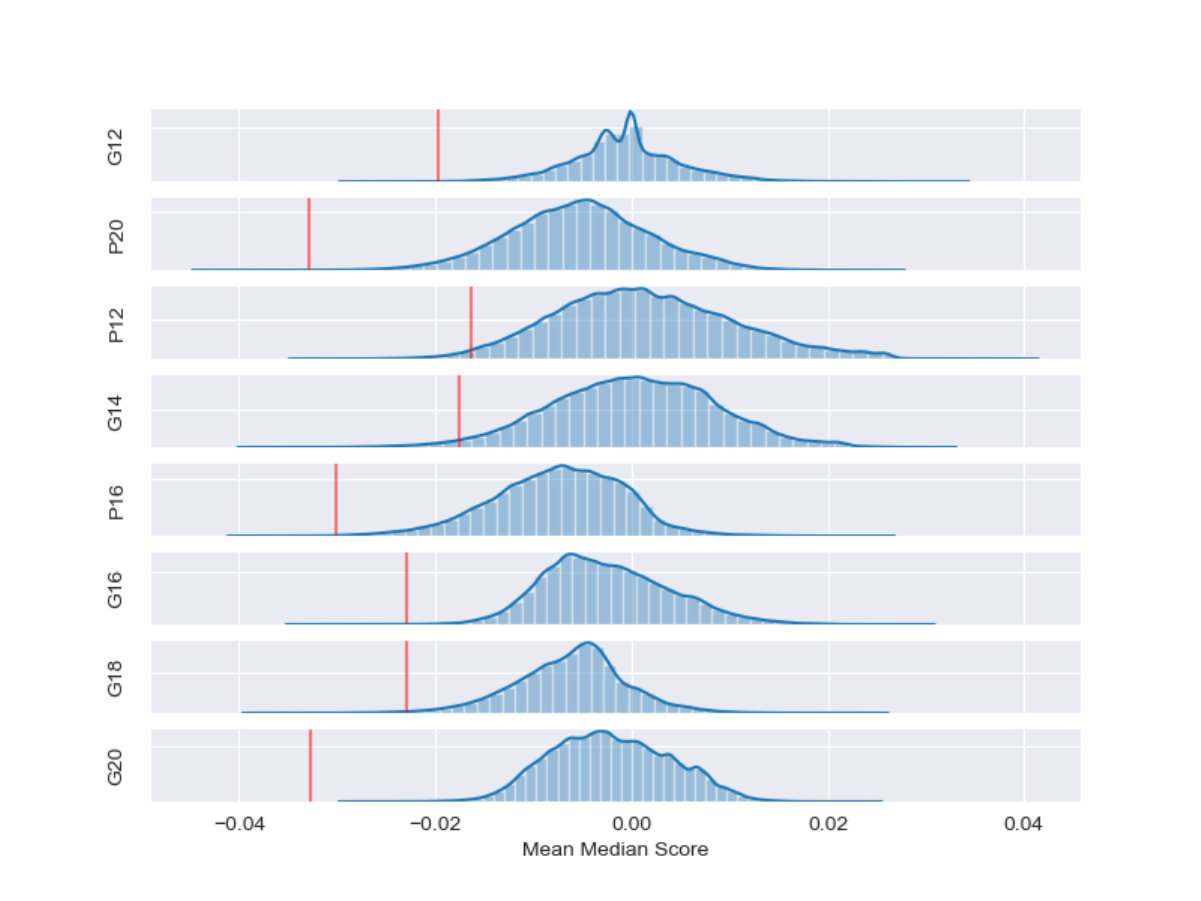}}
    \subfigure[]{\includegraphics[width=0.49\linewidth]{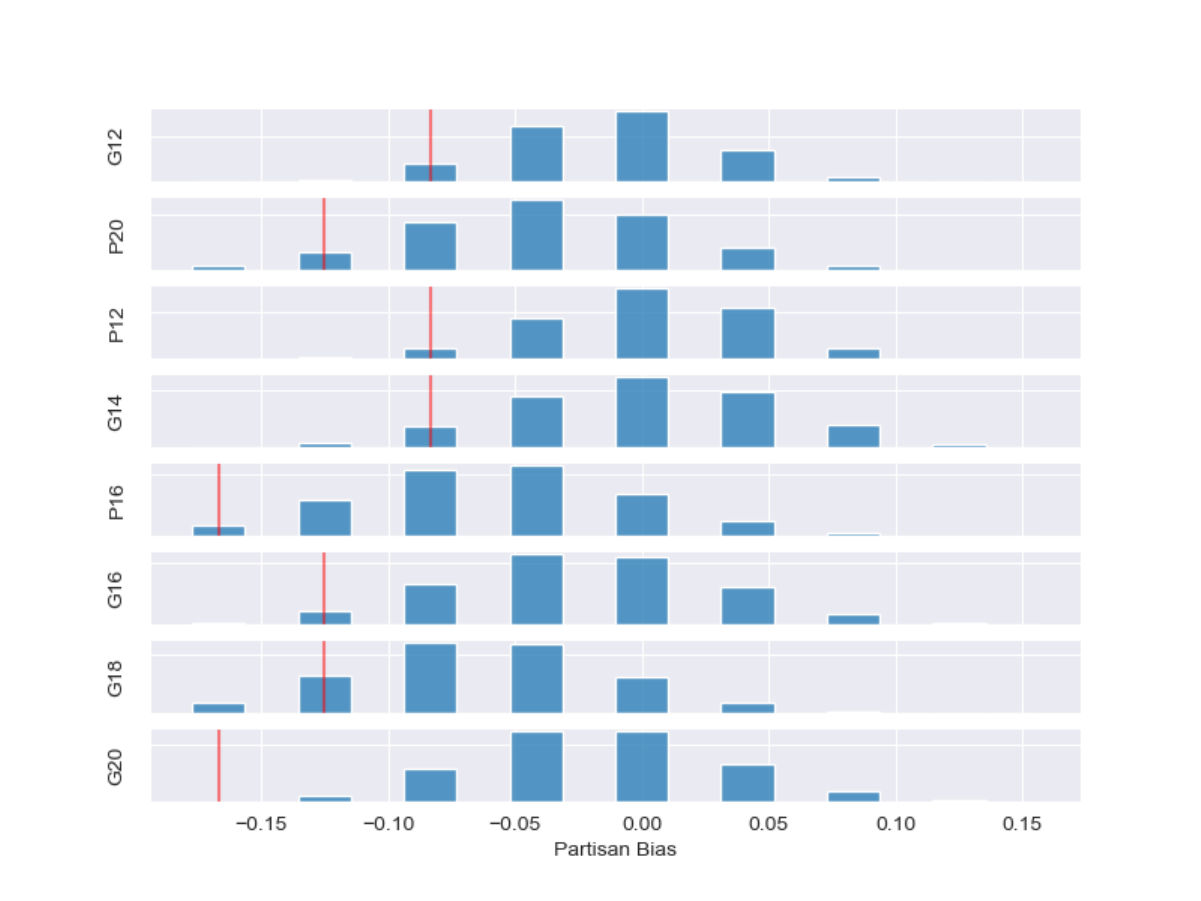}}
    \caption{State Senate Mean Median Scores (a) and Partisan Bias Scores (b) across elections}
    \label{fig:ss-mm}
\end{figure}

This conclusion is supported by the mean-median scores. In Figure \ref{fig:ss-mm}(a), the enacted plan lies on the Republican-leaning tail of the ensemble across elections. The partisan bias is consistent with the mean-median score; the enacted plan is consistently Republican-leaning and outside the typical range, as illustrated in Figure \ref{fig:ss-mm}(b).  

\begin{figure}[!htbp]
    \centering
    \includegraphics[scale=0.5]{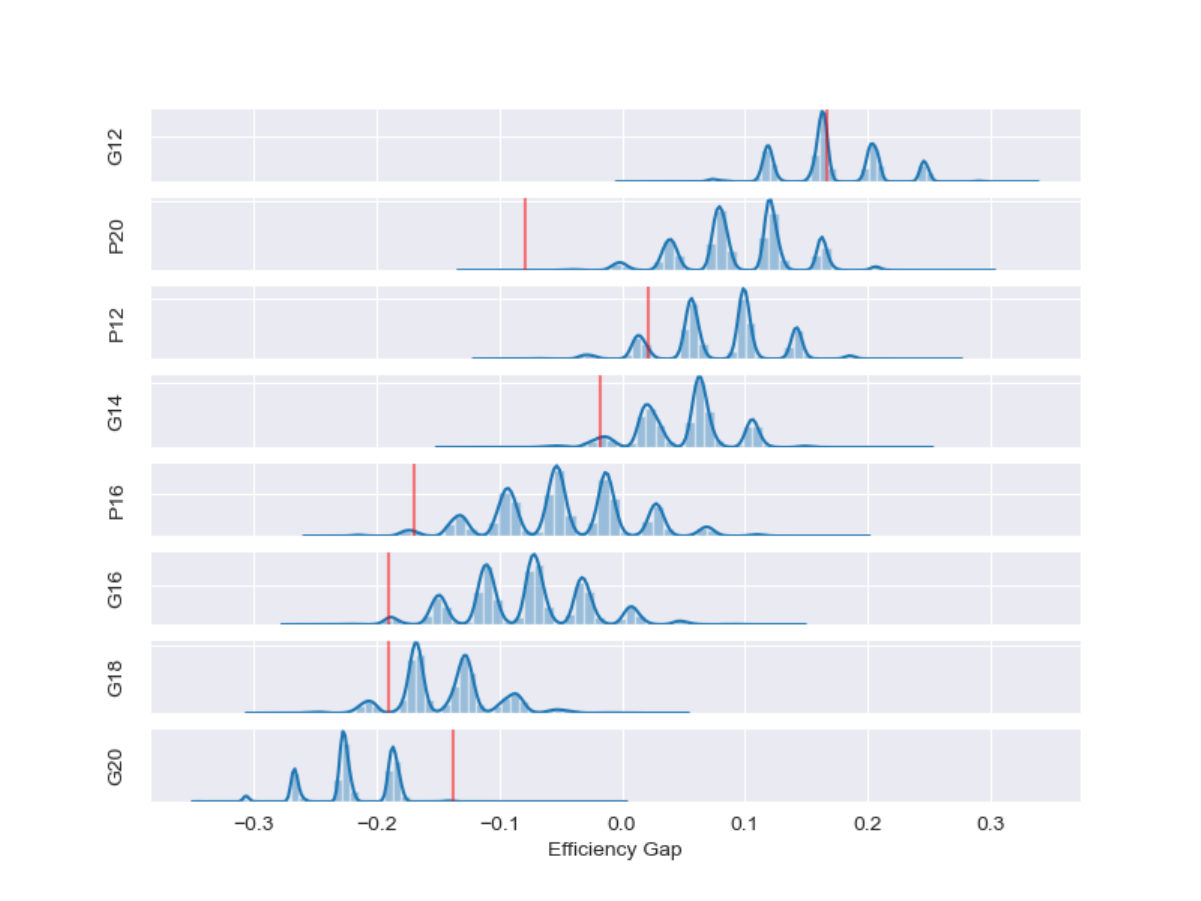}
    \caption{State Senate Efficiency Gap Across Elections}
    \label{fig:ss_eg}
\end{figure}

The efficiency gap, plotted in Figure \ref{fig:ss_eg}, shows a similar result to the Executive council. In close elections, the enacted plan lies on the Republican-leaning edge of the distribution, and in elections won by large margins, the enacted plan falls in the middle or Democrat-leaning side of the distribution.

We conclude that the enacted plan encodes a greater Republican bias than the ensemble, across a wide range of election data and metrics.

\section{Conclusions and Further Work}

The analysis presented here supports the argument that the New Hampshire State Senate and Executive Council maps may have been drawn with partisan interests, as the values of the enacted plans are often outliers compared to those observed on the ensemble. In the Executive Council, partisan symmetry measures point towards a Republican advantage. Investigation of typical vote totals of Hanover, Keene, and Concord suggest packing in district two. A similar results holds for the State Senate, where the enacted plan is a Republican leaning outlier on measures of partisan symmetry.  These results hold even accounting for configurations of specific cities which lead to bimodality in the vote distributions. 

Our result validates some of the claims made in Brown v. Scanlan \cite{complaintNH}. In the Executive Council there is a pattern of packing Democrats into one district to increase Republican representation in the rest of the state. In the State Senate, Democratic representation in the middle districts is lower than is typical for the state; in close elections this would like result in disproportionately large wins for Republicans. 

The ensemble developed for this work has implications that extend beyond the present analysis. By establishing a baseline of `typical' maps for a given state, generated through a nonpartisan process, the ensemble provides a standard against which any possible map can be measured. This approach not only helps researchers identify when a map has gerrymandered (as in this work) but also promotes the creation of fairer maps. Policy-makers and courts can create and analyze maps in the context of neutral, ensemble-created benchmarks, essentially pre-checking maps for partisanship. In this way the ensemble serves a dual purpose; it can be used as a diagnostic tool to detect partisan gerrymandering while also promoting maps that perform well without relying on complicated and often ill-defined measures. 

In sampling from these spaces, novel techniques for metropolizing the recombination algorithm \cite{autry2023metropolized, autry2021metropolized} could give researchers the ability to target a specific distribution, which is an improvement on the current method as the target distribution is unknown \cite{cannon2022spanning}. By explicitly setting a target distribution, researchers can investigate the tradeoffs and effects of different redistricting criteria. For example, one could investigate whether strict enforcement of municipal boundaries benefits Republicans by packing Democratic voters into a small number of high-margin districts, a phenomenon observed in some analyses of partisan bias \cite{chen_unintentional_2013}. Additionally, while progress has been made in sampling single-member districts, there remains a gap in sampling from multi-member and floterial districts like those of the New Hampshire State Congress. Addressing these gaps can further bridge the gap between theoretical redistricting research and practical policy applications.

\begin{appendices}

\section{Mixing Heuristics for Markov Chains}\label{secA1}
The following tables present results for our mixing diagnostics for the Markov chains used to create our ensembles. Two key heuristics were used to assess mixing: the autocorrelation decay and the pairwise Kolmogorov-Smirnov (KS) distance between chains started at different initial states. 

Tables \ref{tab:ss_ac} and \ref{tab:ec_ac} contain autocorrelation lag times. We compute autocorrelation for very long runs of the chain (20 million for the State Senate and 5 million for the Executive Council) to determine the step at which the autocorrelation drops below 0.01. These values are reported in the tables. This value provides an estimate of how long it takes for the chain to `forget' its past. 

Tables \ref{tab: ss_ks} and \ref{tab:ec_ks} contain pairwise KS distances. To verify convergence from different starting points, 10 chains are started from random initializations, and after many steps (10 million for the State Senate and 500,000 for the Executive Council) pairwise KS distances are calculated between each chain. The average of these distances is reported in the tables. When a chain mixes well, it converges to the same distribution regardless of starting point, and therefore average pairwise KS distance is low. 

These diagnostics support our choice of chain lengths by demonstrating that the chains have sufficiently mixed under both heuristics. 

\begin{table}[ht]
    \centering
    \begin{tabular}{llllll}
    \hline
    & 1 & 2 & 3 & 4 & 5\\
    \hline
        GOV12 &  0.00421591 & 0.00740787 & 0.00592862 & 0.0056296 & 0.00729267 \cr
        GOV14 &  0.00521844 & 0.00735707 & 0.00549542 & 0.0057540 & 0.00831342 \cr
        GOV16 &  0.00481853 & 0.00638791 & 0.00508871 & 0.00571370 & 0.00803164 \cr
        GOV18 &  0.0052032  & 0.00770453 & 0.00496964 & 0.00530796 & 0.00781342 \cr
        GOV20 &  0.00431796 & 0.00693604 & 0.00498222 & 0.00588907 & 0.00838267 \cr
        PRES12 & 0.00426382 & 0.00586973 & 0.00512138 & 0.00611364 & 0.00789938 \cr
        PRES16 & 0.00602307 & 0.00729373 & 0.0047844  & 0.00586324 & 0.00732449 \cr
        PRES20 & 0.00691218 & 0.0071284  & 0.00433191 & 0.00664844 & 0.0071012\\
    \hline
    \end{tabular}
    \caption{Average Pairwise KS statistics for the 5 Executive Council Districts, sorted from least to greatest Republican vote total}
    \label{tab:ec_ks}
\end{table}

\begin{table}[ht]
    \centering
    \begin{tabular}{llllll}
    \hline
    & 1 & 2 & 3 & 4 & 5\\
    \hline
        GOV12 & 114.4 & 154.3 & 128.5 & 126.9 & 125.1 \cr
        GOV14 & 94.9  & 158.7 & 143.4 & 122.5 & 170.3 \cr
        GOV16 & 94.4  & 126.9 & 123.8 & 110.0 & 144.3 \cr
        GOV18 & 116.4 & 143.2 & 114.4 & 127.0 & 139.3 \cr
        GOV20 & 128.0 & 124.0 & 124.5 & 101.5 & 130.4 \cr
        PRES12 & 116.1 & 146.5 & 146.5 & 108.2 & 139.3 \cr
        PRES16 & 125.1 & 158.1 & 113.6 & 143.3 & 144.5 \cr
        PRES20 & 145.9 & 153.7 & 85.0  & 163.4 & 144.0\\
    \hline
    \end{tabular}
    \caption{Average lag to attain autocorrelation below 0.01 for sorted Republican vote totals by district}
    \label{tab:ec_ac}
\end{table}

\begin{landscape}
\begin{table}[ht]
\centering
\begin{adjustbox}{width=1\columnwidth}
\small
\begin{tabular}{lrrrrrrrrrrrrrrrrrrrrrrrr}
  \hline
 & 1 & 2 & 3 & 4 & 5 & 6 & 7 & 8 & 9 & 10 & 11 & 12 & 13 & 14 & 15 & 16 & 17 & 18 & 19 & 20 & 21 & 22 & 23 & 24 \\ 
  \hline
GOV12 & 1238 & 2106 & 1938 & 2266 & 5943 & 598334 & 619214 & 359260
  & 81182 & 3973 & 3548 & 18432 & 299427 & 157038 & 140101 & 107504 & 6353 & 6550 & 481545 & 846727 & 739880 & 260004 & 45822 & 217574 \\ 
  GOV14 & 1338 & 2024 &  2230  & 2388  & 3761 &  2091 & 2709 &  3756 &  21275 & 587841 & 672552 &  737484 & 708065 & 401450 & 121412 &  3096 &  633752 & 378601 & 304930 & 652298 & 805534 & 572029 & 665801 & 629772\\ 
    GOV16 & 1984  & 1986  & 2418 &  2185 &144573 &633151 &516985 &122764&
  45728  & 84260 &121140& 152219& 118631 & 74467 &118445 & 14891
  &66633 & 220771 &582556 &691532 &750305 &777262 &493579 &232332\\ 
  GOV18 & 1493 &  2163  & 2249   & 4800 &359145 &711639 &142440  & 3772&
   3350  & 3156  & 3297 &  3641 & 25259 &  16811 & 17823  & 5935&
  67646 & 698479& 823914 &758746& 151013  & 7507 &  9516 &  6266\\ 
  GOV20 & 2017  & 2610  & 2194  &  2530& 228560 &663307 &119787  & 4718&
   3708 &  6162 &219228& 358394 &342834& 272790 & 71890  &  9434&
  81905 &119103 &545861 &751809& 774981& 781654 & 24042 &247420\\ 
  PRES12 & 1224  & 1950  & 1821 &   2237 & 104879 &293840 &283439 &414518&
 579565 &580611 &428415 &227647&  32917  &19806 &125947 &134463&
   9619 & 16255 &514178 &824383 &840295 &190794& 394800 &221026 \\ 
  PRES16 & 2535  & 2168  & 5279  & 8925 &268173 &466493  & 8651 &  4563&
 102687 &373336 &490866 &252351  &34355 &  6558  & 3592 &   5716&
 239417& 573644 & 689886& 628293 &539205 &157697  &10662 & 72353\\ 
  PRES20 & 2958  & 2304  & 5198 &  17602 &479412 &542392  & 6468 & 67669&
 567680  &638801 &597633 &181521  & 5180 &  4562&  56705 &285757&
 407353 &606716 &625285 &420508 &335653 & 29033&  7357 & 12449 \\ 
   \hline
\end{tabular}
\end{adjustbox}
\caption{Average lag to attain autocorrelation below 0.01 for sorted Republican vote totals by district} 
\label{tab:ss_ac}
\end{table} 

\begin{table}[h]
\centering
\begin{adjustbox}{width=1\columnwidth}
\small
\begin{tabular}{lrrrrrrrrrrrrrrrrrrrrrrrr}
  \hline
 & 1 & 2 & 3 & 4 & 5 & 6 & 7 & 8 & 9 & 10 & 11 & 12 & 13 & 14 & 15 & 16 & 17 & 18 & 19 & 20 & 21 & 22 & 23 & 24 \\ 
  \hline
GOV12 & 0.00746421 & 0.00738492 & 0.0059108  & 0.00820326 & 0.03166541 & 0.05272608
 & 0.0690287  & 0.04742722 & 0.03170964 & 0.01596767 & 0.01942463 & 0.03570441
 & 0.05301232 & 0.04947304 & 0.04840355 & 0.04973123 & 0.02962903 & 0.02415266
 & 0.07593755 & 0.12276848 & 0.12152078 & 0.04807615 & 0.06103478 & 0.09263022\\
GOV14 & 0.00629292 & 0.00561628 & 0.00661108 & 0.00727118 & 0.00979322 & 0.0064813
 & 0.01618599 & 0.0162338  & 0.03287922 & 0.04952341 & 0.07450262 & 0.08116674
 & 0.07982718 & 0.05546073 & 0.02692621 & 0.03167879 & 0.08009767 & 0.06322768
 & 0.04422766 & 0.08380088 & 0.12140145 & 0.10255346 & 0.09821423 & 0.15932963\\
GOV16 & 0.00860536 & 0.00723338 & 0.00757485 & 0.02500664 & 0.05643081 & 0.05181119
 & 0.07440228 & 0.03565294 & 0.02259418 & 0.02169706 & 0.02755694 & 0.03160729
 & 0.04655741 & 0.06075286 & 0.04683201 & 0.03238795 & 0.0403495  & 0.04272124
 & 0.05615057 & 0.10367346 & 0.14655651 & 0.09847478 & 0.08245241 & 0.14626076\\
GOV18  & 0.00772433 & 0.00576409 & 0.00741841 & 0.02082957 & 0.04769098 & 0.09432386
 & 0.04148392 & 0.01133091 & 0.01120564 & 0.02078481 & 0.03685532 & 0.04873831
 & 0.03918144 & 0.0406977  & 0.03893848 & 0.02969813 & 0.03915364 & 0.08377632
 & 0.10849367 & 0.11246157 & 0.05638986 & 0.02656568 & 0.04616468 & 0.08091433\\
GOV20  & 0.00839746 & 0.00718508 & 0.00654122 & 0.01392847 & 0.04483915 & 0.07866748
 & 0.04700421 & 0.01478506 & 0.00828353 & 0.01262386 & 0.02959439 & 0.04910098
 & 0.05721086 & 0.06578272 & 0.04671448 & 0.02765584 & 0.03888515 & 0.0423092
 & 0.06207852 & 0.10474533 & 0.09647605 & 0.11799778 & 0.05910955 & 0.14626076\\
PRES12   & 0.00657057 & 0.0072771  & 0.00607774 & 0.02107175 & 0.05331691 & 0.04130823
 & 0.03011342 & 0.04708978 & 0.0597832  & 0.05462279 & 0.05257504 & 0.0374573
 & 0.03315634 & 0.03846967 & 0.04630505 & 0.05145654 & 0.04609867 & 0.0285254
 & 0.0542849  & 0.12423302 & 0.12767322 & 0.0972663  & 0.07392876 & 0.0925786 \\
PRES16   & 0.00867873 & 0.00933213 & 0.01049136 & 0.0286223  & 0.04620154 & 0.06422188
 & 0.03704703 & 0.01846614 & 0.02977595 & 0.04283395 & 0.06101838 & 0.06619397
 & 0.05005875 & 0.0278989  & 0.02420644 & 0.02499964 & 0.03618181 & 0.05087962
 & 0.11576503 & 0.1036124  & 0.07112142 & 0.0616693  & 0.05927408 & 0.07916322\\
PRES20 & 0.00776663 & 0.00808934 & 0.01082212 & 0.03317619 & 0.05978789 & 0.0755572
 & 0.03007401 & 0.03696558 & 0.05336268 & 0.05729177 & 0.0632198  & 0.04931742
 & 0.03536015 & 0.02968275 & 0.03414217 & 0.03657057 & 0.03742854 & 0.06358969
 & 0.1248624  & 0.06893066 & 0.06500145 & 0.0503125  & 0.0476214  & 0.05822621\\
   \hline
\end{tabular}
\end{adjustbox}
\caption{Average Pairwise KS statistics for the 24 State Senate Districts, sorted from least to greatest Republican vote total} 
\label{tab: ss_ks}
\end{table} 
\end{landscape}
\end{appendices}

\bibliography{Gerry_refs}

\end{document}